# About the Nature of the Quantum System – an Examination of the Random Discontinuous Motion Interpretation of the Quantum Mechanics

*by Sofia D. Wechsler*

**Abstract**
What is the quantum system? Consider the wave-function of the electron – what we call 'single particle wave-function' – and assume that it contains *N* wave-packets. If we pass all the wave-packets through an electric field, all are deflected, as if each one of them contains an electron. However, if we bring any two wave-packets to travel close to one another, they don't repel one another, as if at least one of them contains no charge.
In trying to solve the measurement problem of the quantum mechanics (QM), different interpretations were proposed, each one coming with a particular ontology. However, only one interpretation paid explicit attention to the contradiction mentioned above. This interpretation was proposed by S. Gao who named it 'random discontinuous motion' (RDM), because it assumes the existence of a particle that jumps from place to place at random. The particle carries all the physical properties of the respective type of particle, mass, charge, magnetic momentum, etc. It jumps under the control of an "instantaneous condition" about which Gao did not give details so far.
Along with presenting problems of the QM that this interpretation solves, this text reveals difficulties vis-à-vis entanglements and the special relativity.

**Abbreviations**
dBB     = de Broglie-Bohm
CSL     = continuous spontaneous localization
QM      = quantum mechanics
QS      = quantum system
RDM     = random, discontinuous motion
w-f     = wave-function
w-p     = wave-packet

## 1. Introduction

The measurement problem in the quantum mechanics (QM) lead to different so-called 'interpretations', each one coming with an assumption of what can be the quantum system (QS) from ontological point of view. Here are several wide-spread interpretations and their main assumptions:

- de Broglie and Bohm (dBB) [1, 2] assumed the existence of a particle traveling on a trajectory guided by the wave-function (w-f), considered a reality;
- the 'full and empty waves' hypotheses [3] assumed that the QS consists in two types of waves, one able to impress a detector (full wave) and one unable to do so (empty wave);
- the 'consistent histories' [4] admitted that the QS travels at once on two paths, though without being simultaneously on both;

- the 'transactional' interpretation [5] suggests that the source of QSs emits waves traveling forward in time, and the detectors emit waves traveling backward in time. A 'hand-shake' is postulated between a wave from the source and the wave from one of the detectors, determining which detector would click;
- Everett's or 'many-worlds' interpretation [6] supposes the existence of parallel worlds, each one equipped with detectors. Each wave-packet (w-p) of the w-f triggers a detector, though in another world.
- Ghirardi, Rimini, Weber and Pearle accepted the collapse of the w-f as a real phenomenon, and proposed to modify the Schrödinger equation such as its solution become localized into a very small region in space [7, 8, 9]. The most advanced form of this proposal – the CSL model of collapse – is examined in detail in [10].

These interpretations are discussed in [11], where it is explained that each one of [1, 2] thru [5] introduced a change in the QM formalism, and this change lead to predictions which disagree with the QM. About [6], besides the fact that the general relativity theory provides no ground to support the existence of additional worlds, an oddity is shown for which this interpretation has no explanation.

Another interpretation is due to S. Gao [12, 13, 14], and will be examined in detail in this work. Gao called attention upon an experimental fact, namely, any two parts of the w-f of an electron don't repel one another.

> "If the wave function represents a physical field, then it seems odd that there are (electromagnetic and gravitational) interactions between the fields of two electrons but no interactions between two parts of the field of an electron."

In [14] Gao exemplified the non-self-interaction showing that if an electron w-f consists in two w-ps, $|\psi\rangle_A$ and $|\psi\rangle_B$, they do not repel one another. That means, $|\psi\rangle_A$ does not feel the existence of $|\psi\rangle_B$ as if $|\psi\rangle_B$ is empty, and vice-versa. Starting from this phenomenon, Gao proposed an interpretation of the QM built on the exclusive ontology of *particles*. A particle is supposed to be in random discontinuous motion (RDM), jumping from one position to another.

> "the superposition principle of quantum mechanics requires that the charge distribution of a quantum system such as an electron is effective; at every instant there is only a localized particle with the total charge of the system, while during an infinitesimal time interval around the instant the ergodic motion of the particle forms the effective charge distribution at the instant" [12]

The present article adds arguments in favor of the RDM interpretation, and describes two experiments from which one was already performed, to test them. However, it is also argued that RDM has many weak points.

**Note**: in this article the expression 'quantum particle' is frequently used. It may mean an elementary particle, or an atom or molecule for which the internal structure is ignored. In any case, the respective item is considered as described by the QM, not be the classical physics.

The rest of the article has the following content: section 2 exposes phenomenological constraints which stand at the base of the RDM interpretation. It is shown that these constraints are incompatible, that they lead to a contradiction. Section 3 presents two experiments which justify the constraints. Section 4 points to weaknesses of the RDM. Section 5 contains conclusions.

## 2. Constraints on the wave-function of a quantum particle

Three phenomenological facts stand at the base of the RDM.

*1. Non-self-interaction.*

This feature was mentioned in the previous section, and now it will be exemplified. Consider the Hamiltonian of an electron in the hydrogen atom,

$$\hat{H}(\mathbf{r}) = \frac{\hat{\mathbf{P}}^2}{2\mu} + U(\mathbf{r}), \tag{1}$$

where $\hat{\mathbf{P}}$ is the linear momentum operator, $\mu$ is the electron reduced mass, $\mathbf{r}$ is a position with respect to the nucleus and $U(\mathbf{r})$ is the potential energy of an electron at this position.

Let's consider the orbital of the electron in the ground state. According to the standard QM the position of the electron inside the cloud is undefined. However, the Hamiltonian shows that if there is a charge at the position $\mathbf{r}$, the only field acting on it is the electrostatic field between that charge and the nucleus. The Hamiltonian contains no term as $U_1(\mathbf{r},\mathbf{r}')$ where $\mathbf{r}$ and $\mathbf{r}'$ are two positions at which the electron charge is present simultaneously. So, if there is an electron charge at some position $\mathbf{r}$, the rest of the cloud does not act on it. What one can infer from this is that the rest of the cloud is chargeless, and the charge at $\mathbf{r}$ is the entire electron charge, $-e$.

*2. Presence of all the w-ps in each trial of the experiment.*

If the w-f of a quantum particle contains $N$ w-ps, in each trial and trial of an experiment all the w-ps are present. This property is a simple generalization of the case with two w-ps, where if in each trial of the experiment only one or the other of the w-ps is present, in the end of the experiment would appear on the screen instead of fringes, two spots. With $N$ w-ps, if in each trial of the experiment only one or another of the w-ps is present, the image resulting on the screen in the end of the experiment would be $N$ spots, no fringes.

*3. Multiplication of physical properties.*

Consider the w-f of a quantum particle carrying a certain charge, and consisting in $N$ w-ps. If each w-p passes through a field which acts on the respective charge, each w-p is deflected as if it carries the entire charge. This feature is not obvious so far, and will be proved in the following sections.

**Inference:**

These three features impose opposite constraints when trying to build an ontology for interpreting the QM.

Indeed, if each w-p is present in each trial of an experiment and hosts all the physical properties of the respective type of experiment in entirety, the w-f of an electron with $N$ w-ps should carry $N$ charges. The beam-splitter of the w-f of a charged particle seems to be a charge generator. That's obviously absurd.

It seems that the only solution is to suppose the existence of an object that carries all these properties, and which jumps from w-p to w-p all the time during the trial. In this way, two w-ps of the w-f of a charged particle won't repel one another, because they are not simultaneously charged. With this idea, the probability that a w-p trigger a detector would be proportional with the frequency with which the 'properties carrier' lands in that w-p. Regrettably, the idea of a jumping object is not flawless, as will be shown later in this text.

## 3. Two experiments

The experiments described in this section justify the constraints exposed in the previous section – especially the constraint 3. The description is in the 2D geometry.

## 3.A. An electron in electric field

A single-particle w-p of an electron lands on an imbalanced beam-splitter BS – figure 1. Each w-p passes through a capacitor with the field perpendicular to the axis **z**. Each capacitor is the mirror of the other, so, the electric fields are of the same intensity, but opposite in direction. For a sufficiently big intensity of the field, the w-ps are deflected enough strongly so as to meet one another on a sensitive plate S.

The electron state after exiting the beam-splitter BS is

$$\psi(\mathbf{r},t) = \alpha\psi_1(\mathbf{r},t) + i\beta\psi_2(\mathbf{r},t), \tag{2}$$

where $\alpha^2 + \beta^2 = 1$, and we take $\alpha$ and $\beta$ as real. The Hamiltonians in the capacitors are,

$$\hat{H} = \hat{H}_0 \pm eEx, \tag{3}$$

where $\hat{H}_0$ is the Hamiltonian of the electron in the free space, and the sign '+' ('–') is for $C_2$ ($C_1$).
Using the potential energy from (3) in Ehrenfest's theorem,

$$d\langle p_x \rangle / dt = -\langle dV/dx \rangle, \tag{4}$$

then, integrating the equation (4) by time, one finds that the change in the average x-component of the linear momentum inside the capacitors is

$$\Delta\langle p_x \rangle = \pm eE\Delta t, \tag{5}$$

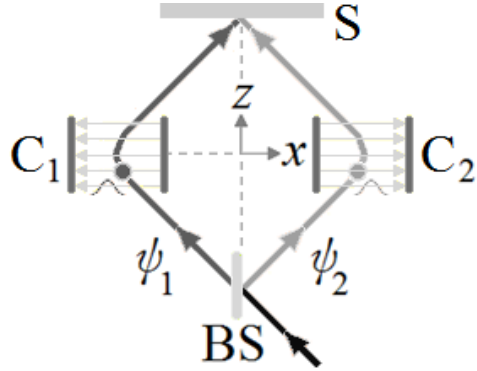

Figure 1. An experiment with electrons.
A single-particle w-p of an electron lands on a non-balanced beam-splitter BS. The transmission coefficient is depicted as greater than the reflection coefficient so that the intensity of the w-p $\psi_1$ is greater than that of $\psi_2$. BS is aligned to the axis **z**. $C_1$ and $C_2$ are charged capacitors, each one being the mirror of the other. The electric fields are parallel to the axis **x**, though opposite in direction, so, the motion of the w-ps along the axis **z** is not affected. The dots inside the capacitors represent the w-ps. S is a sensitive screen.

where $\Delta t$ is the interval of time spent in the capacitors, and '+' ('–') is for $C_1$ ($C_2$). Thus, $\psi_1$ is deflected in the direction **x**, while $\psi_2$ is deflected in the direction –**x**.

What is relevant for the present analysis is the pattern formed on the screen. For this reason we restrict the analysis below to the evolution is the x-direction. On the screen surface is convenient to express the w-ps as

$$\psi_1(x, z_S, t) = \mathcal{N} G_1(x, t) e^{iP_x x/\hbar},$$
$$\psi_2(x, z_S, t) = \mathcal{N} G_2(x, t) e^{-iP_x x/\hbar}. \tag{6}$$

$z_S$ is the screen height, $\mathcal{N}$ a normalization factor, $P_x$ the average **x**-component of the linear momentum of $\psi_1$. We assume the functions $G$ as Gaussians quite flat in the direction **x**. Due to the symmetry with respect to the axis **z**, $G_1(x, t) = G_2(-x, t)$. Thus, at the screen, the w-f will be

$$\psi(x, z_S, t) = \mathcal{N} \left[ \alpha G_1(x, t) e^{iP_x x/\hbar} + i\beta G_1(-x, t) e^{-iP_x x/\hbar} \right]. \tag{7}$$

After many trials of the experiment, on the screen will appear a pattern of intensity

$$I(x, t) = \mathcal{N}^2 \left[ \alpha^2 G_1^2(x, t) + \beta^2 G_1^2(-x, t) + 2\alpha\beta G_1(x, t) G_1(-x, t) \sin\left(\frac{2P_x x}{\hbar}\right) \right]. \tag{8}$$

The flatness of the Gaussians, allows using the approximation

$$I(x, t) = \mathcal{N}^2 G_1^2 \left[ 1 + 2\alpha\beta \sin(2P_x x/\hbar) \right], \tag{9}$$

which is an interference pattern superimposed on a ground constant for a while during the trial.

**Implications:**
according to the QM one concludes the following:

a) The interference fringes prove that both w-ps are present in each trial and trial of the experiment. This implication may be immediately generalized to the case of a w-f with several w-ps. For instance, interference can be obtained also with four w-ps, though the pattern would be more complicated.

b) As the Hamiltonian (3) shows, each w-p carries the entire elementary charge, while the absolute square of the amplitude of the w-p, in our case $\alpha^2$ or $\beta^2$, has no relevance. Experimentally, the charge can be inferred from a graph $\Delta\langle p_x \rangle / (E \Delta t)$ drawn for various values of $E$, $\Delta t$, and corresponding obtained deflections $\Delta\langle p_x \rangle$ – equation (5).

The implication (*a*) justifies the constraint (*2*) from the previous section. The implication (*b*) justifies the constraint (*3*). In particular, it also rules out a proposal due to Bedingham et al. [15], by which a quantum particle is an object distributed in space.

## 3.B. The Aharonov-Casher experiment

After the discovery and experimental test of the Aharonov-Bohm effect [16, 17, 18, 19], its dual, known as the Aharonov-Casher effect, was discovered and implemented experimentally [20, 21, 22, 23, 24].

Using this effect, the present subsection exemplifies the constraint (*3*) on another physical property than the electrical charge, the magnetic moment of a neutron. The experiment is taken from the report [22]. Figure 2 below, illustrates the main components of the configuration. According to the report, the experiment was carried with non-polarized neutrons; though for simplicity, in this text the neutrons will be assumed polarized perpendicularly to the figure.

A beam of neutrons falls on the planes S of a perfect Si crystal. Here, the incident beam undergoes Bragg diffraction and a transmitted and a reflected wave are obtained. Each wave passes through the electric field created between two electrodes oppositely charged. Then, they illuminate another array of crystal planes M. These planes reflect the waves towards a third array of crystal planes, A, where interference occurs. The neutrons exiting the array A are detected by two detectors $C_2$ and $C_3$.

As simply explained in [22], the canonical momentum of a neutron of mass $m$, magnetic moment $\mathbf{\mu}$, and velocity $\mathbf{v}$, in an electric field $\mathbf{E}$, is

$$\mathbf{p} = m\mathbf{v} + \frac{\mathbf{\mu} \times \mathbf{E}}{c^2}.^1 \qquad (10)$$

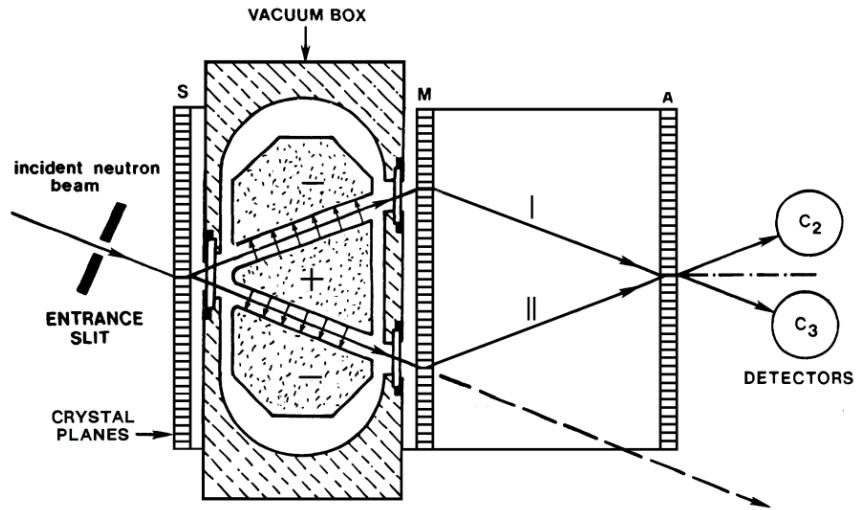

Figure 2. The Aharonov-Casher experiment implemented with neutrons.
This figure in adapted from the figure 2 in [22]. Since the discussion here is rather conceptual the numerical and some of technical details were ignored. A beam of neutrons is split into a transmitted and a reflected part on the planes S of a perfect Si crystal. Each part passes through a channel between a prism-shaped electrode positively charged, and a polygonal electrode negatively charged. The small arrows indicate the direction of the electric field. Another array M of crystal planes reflects the beams toward a third array of planes A, where the two beams meet and interfere. The detectors $C_2$ and $C_3$ collect the neutrons exiting the array A.

---

[1] According to [20], in (10) should appear only c instead of $c^2$. However, the consistence between the different parts of the formula in the units of measurement, requires $c^2$.

The difference between the phase acquired by the lower w-p during the travel from the array S to the array A, and the phase acquired by the upper w-p during the travel between these arrays, is equal to the line integral over **p** along the closed loop around the prism electrode, in counter-clock direction,

$$\Delta\varphi = \frac{1}{\hbar}\oint \mathbf{p}\,d\mathbf{r} = \frac{1}{\hbar c^2}\oint \mathbf{\mu}\times\mathbf{E}\,d\mathbf{r}. \tag{11}$$

With the magnetic moment **μ** perpendicular on the plane of the figure and pointing upwards, the vector product $\mathbf{\mu}\times\mathbf{E}$ is parallel (anti-parallel) with the velocity of the w-p in the lower (upper) electric field.

What is relevant for the present analysis is the fact that each one of the two wave-packets carries the magnetic moment **μ**, as says the constraint (*3*) from the section 2.

## 4. Criticism of the random discontinuous jumps interpretation

This section points to problems still to be solved by the RDM interpretation of the QM.

*i) A controller is needed.*
Assume that the w-f of a quantum particle is prepared in the Gaussian form, and is let to propagate in the free space. It is known that the Gaussian expands during the time, though maintains the Gaussian form. According to the RDM interpretation, the particle which carries the physical properties jumps all the time from position to position. If the jumps of the particle are really at random it is not clear how the Gaussian form of the wave-packet is preserved.
In general, the interval of time the particle is present in a w-p has to be proportional with the absolute square of the amplitude of that w-p. It's obvious that an additional element, connected to the w-f has to be added to the RDM ontology for controlling where and for how much time may the particle sojourn.

Let's consider now a single particle w-f with two w-ps, $\psi_1$ and $\psi_2$, and let's introduce a detector $D_1$ on the path of $\psi_1$, and detectors $D_2$ and $D_3$ on the path of $\psi_2$ – $D_3$ follows $D_2$ – figure 3. Let the path to $D_1$ be longer than the path to $D_3$. Assume also that the detectors are ideal. We denote by V the virtual event that the w-p $\psi_2$ meets $D_3$ if $D_2$ would have been removed, and by $t_0$ the corresponding time. By E we will denote the event that the w-p $\psi_1$ reaches some position by this time.
We will concentrate on trials in which $D_2$ remains silent. The RDM interpretation explains this silence by suggesting that during the contact between the w-p $\psi_2$ and $D_2$ the particle was in the w-p $\psi_1$. However, later on, the particle may jump from $\psi_1$ to $\psi_2$ and make $D_3$ click. The experimental practice shows that such a thing never happens. Then it remains to assume that there exist a locking process by which the particle remains in $\psi_1$ for $t \geq t_0$. The RDM has to be supplemented with such a locking process.

*ii) Problems with the relativity.*
Introducing relativistic considerations the case becomes more severe. Let's move the detectors $D_2$ and $D_3$ at different positions closer and closer to the source as depicted in the figure 3 ($D_2$ is not shown in the figure,

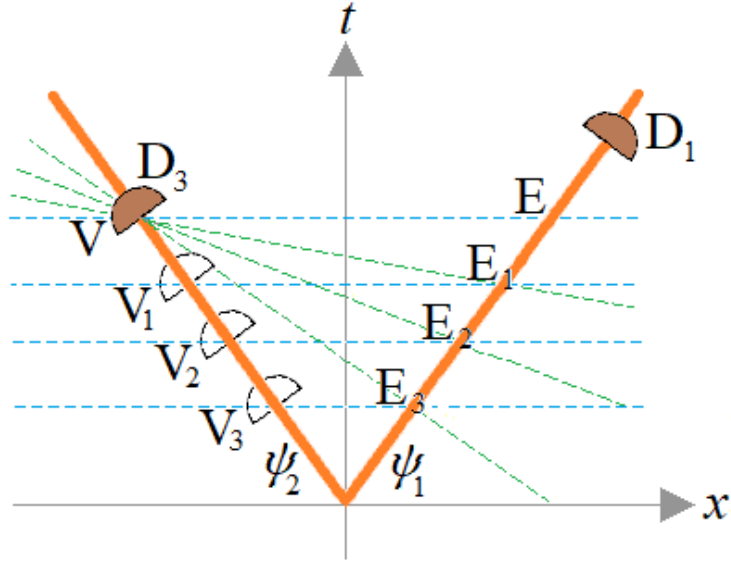

Figure 3. RDM vs. the relativity.

The world lines of two w-ps – thick, orange lines – $\psi_1$ and $\psi_2$, pass through three detectors $D_1$ on the way of $\psi_1$, $D_2$ and $D_3$ on the way of $\psi_2$ ($D_2$ is not shown but it precedes $D_3$ and is very close to it). $D_3$ and $D_2$ are displaced closer and closer to the source of the particles (the source is not shown, neither the instantiations of $D_2$). The virtual impingement of $\psi_2$ on $D_3$ at the different positions of $D_3$ are marked as virtual events, V, $V_1$, $V_2$, etc. By the respective times, the w-p $\psi_1$ also reaches positions closer to the source, and these events are named E, $E_1$, $E_2$, etc . The dashed light-blue lines denote hypersurfaces of constant time according to the lab frame. The dashed green lines denote hypersurfaces of constant time in the frames of coordinates $\mathcal{F}_1$, $\mathcal{F}_2$, etc.

but it precedes $D_3$ very closely). The virtual event V and the simultaneous (by the lab frame) event E, occur sooner and sooner; let's name the new instantiations $V_1$ and $E_1$. If the relativistic interval between the events V and $E_1$ is space-like, there exists a frame of coordinates $\mathcal{F}_1$ by which V and $E_1$ are simultaneous.

If the relativistic intervals between V and $E_1$, $E_2$, etc. are still space-like, precede V, there exist frames, $\mathcal{F}_2$, $\mathcal{F}_3$, etc., by which the event V is simultaneous with $E_2$, $E_3$, etc. In the figure are shown only several instantiations of the pairs $V_n$, $E_n$, but in fact they form two continuums.

Returning to the lab frame, since all the events $E_n$ precedes V, it turns out that the particle was locked in $\psi_1$ all the time before V, and all the more before $\psi_2$ met $D_2$, i.e. before any perturbation of the w-f occurred. However, that contradicts the constraint (*2*) and (*3*) from the section 2, and proved by the experiments in the section 3. These constraints say that in each trial of the experiment both w-ps have to be present and both have to be endowed with all the physical properties of the respective type of particle.

That would mean that the particle should be present continuously in both w-ps and never leave any one of them. But there is only one particle.

*iii) A memory is needed.*

If the two w-ps of a w-f are brought to cross one another, in the interference region appear fringes. The pattern will depend on the relative phase between the w-ps. A memory is needed which can recall at each instant the phases acquired by each one of the w-ps.

*iv) The connection with the w-f is not clear*

The connection between the jumps of the particle and the w-f is explained by Gao as follows

> "when the probability density that the particle appears in each position is equal to the modulus squared of its wave function there at every instant, the discontinuous motion will be ergodic and can generate the right charge distribution, for which the charge density in each position is proportional to the modulus squared of its wave function there" [12]

Gao doesn't say what ensures that the probability density be equal to the absolute square of the w-f. But he leaves an open door for the existence of an ontic entity, which he names "instantaneous condition", that makes the particle jump

> "we need to analyze the cause of motion. . . .
> That the instantaneous condition is deterministic means that it leads to a deterministic change of the position of the particle at each instant That the instantaneous condition is indeterministic means that it only determines the probability density that the particle appears in each position in space at each instant."

These inferences are not enough for introducing into the scene the Schrödinger equation. The solution of this equation is more than a set of probabilities, it contains phases, as already said at the entry (*iii*). With a particle jumping at random it is not clear how the forbidden regions would be avoided in a pattern of fringes.

*v) The correlations in the entanglements*

If two particles are entangled, there exist correlations between the w-ps that may respond to tests. For instance, with the polarization singlet there exist sine and cosine law that have to be obeyed along the experiment. It is not clear how that can be done with particles jumping at random.

Gao said in [12] that he intends to work on the instantaneous condition that makes the particles jump, but at present it is not known to the present author where he obtained more precise result that would solve the above problems.

## 5. Conclusions

The RDM interpretation was constructed on the ontology of a particle jumping from place to place under the control of an "instantaneous condition". The purpose of the interpretation was to explain the non-self-interaction property of the QSs, which is a fact experimentally confirmed. None of the other interpretations mentioned in the section 1 pays attention to this phenomenon, except the 'consistent histories' which approaches it only vaguely, and the 'many worlds', which, as pointed out in [11], contains a self-contradiction.

The fact that S. Gao called attention on the non-self-interaction, has an additional important consequence. QM is supposed to predict only what is going to be actually measured. According to the famous dictum of A. Peres "*Unperformed experiments have no result*" [25], nothing more precise than the w-f can be said about a QS, unless a macroscopic test is performed. However, the experimental evidence of the non-interaction of two parts of the w-f is a piece of information on the QS behavior, before the measurement.

The present author's opinion is that the RDM interpretation has to be enhanced with a control process, the "instantaneous condition" that Gao mentioned, process that should solve the problems mentioned in section 4.

The control process should be connected with the wave-function.

——————————————————————

**Acknowledgements**

I am in debt to Prof. Markus Arndt from the Physics Faculty, University of Vienna, for strengthening my conclusion "*Multiplication of physical properties*", (see section 2), and sending me many references which confirm it.

**References**

[1]  L. de Broglie, "*Ondes et mouvements*", publisher Gauthier-Villars, (1926); "*An introduction to the study of the wave mechanics*", translation from French by H. T. Flint, D.Sc, Ph.D., first edition 1930.

[2]  David Bohm, "*A suggested interpretation of the quantum theory in terms of "hidden" variables*", part I and II, Phys. Rev. **85**, pages 166-179, respectively pages 180-193 (1952).

[3]  L. Hardy, "*On the existence of empty waves in quantum theory*", Physics Letters A, vol. **167**, issue 1, pages 11-16 (1992).

[4]  R. J. Griffiths, "*Consistent quantum theory*", Cambridge, U.K.: Cambridge University Press (2002); "*The Consistent Histories Approach to Quantum Mechanics*", Stanford Encyclopedia of Philosophy, (First published Thu Aug 7, 2014; substantive revision Thu Jun 6, 2019).

[5]  J. G. Cramer, "*The Transactional Interpretation of the Quantum Mechanics*", Rev. Mod. Phys. 58, no. 3, pages 647-688, (1986); "*An Overview of the Transactional Interpretation*", Int. J. of Th. Phys. **27**, no. 2, pages 227-236 (1988).

[6]  H. Everett, "*The theory of the Universal Wavefunction*", Thesis, Princeton University, pages 1-140 (1956, 1973).

[7]  G-C. Ghirardi, A. Rimini and T. Weber, "*Unified dynamics for microscopic and macroscopic systems*", Phys. Rev. D **34**, page 470 (1986).

[8]  G-C Ghirardi, P. Pearle, and A. Rimini, "*Markov processes in Hilbert space and continuous spontaneous localization of systems of identical particles*", Phys. Rev. A: Atomic, Molecular, and Optical Physics **42**, no. 1, pages 78-89 (1990).

[9]  A. Bassi and G-C. Ghirardi, "*Dynamical Reduction Models*", Phys. Rept. **379**, page 257 (2003).

[10] S. D. Wechsler, "*In Praise and in Criticism of the Model of Continuous Spontaneous Localization of the Wave-Function*", JQIS **10**, no 4, pages 73-103 (2020).

[11] S. D. Wechsler, "*The quantum mechanics needs the principle of wave-function collapse – but this principle shouldn't be misunderstood* ", JQIS **11**, no 1, pages 42-63 (2021).

[12] S. Gao, "*MEANING OF THE WAVE FUNCTION - In search of the ontology of quantum mechanics*", arXiv:quant-ph/arXiv:1611.02738v1.

[13] S. Gao, "*Is an Electron a Charge Cloud? A Reexamination of Schrödinger's Charge Density Hypothesis*", Foundations of Science **23**, no. 1, pages 145-157 (2018).

[14] S. Gao, "*A puzzle for the field ontologists*", Foundations of Physics **50**, no. 11, pages 1541-1553 (2020).

[15] D. Bedingham, D. Dürr, G.C. Ghirardi, S. Goldstein, and R. Tumulka, "*Matter Density and Relativistic Models of Wave Function Collapse*", Journal of Statistical Physics **154**, pages 623-631 (2014); [arXiv:quant-ph/1111.1425v2].

[16] Y. Aharonov, D. Bohm, "*Significance of Electromagnetic Potentials in the Quantum Theory*", Phys. Rev. **115**, pages 485-491 (1959).

[17] R. G. Chambers, "*Shift of an Electron Interference Pattern by Enclosed Magnetic Flux*", Phys. Rev. Lett. **5**, no. 1, pages 3-5 (1960).

[18] G. Möllenstedt and W. Bayh, "*Messung der kontinuierlichen Phasenschiebung von Elektronenwellen im kraftfeldfreien Raum durch das magnetische vektorpotential einer Luftspule*", Naturwissemchaften **49**, issue 4, pages 81-82 (1962); English translation "*Measurement of the continuous phase shift of electron waves in a force-field-free space through the magnetic vector potential of an air-core coil*".

[19] A. Tonomura, N. Osakabe, T. Matsuda, T. Kawasaki, J. Endo, S. Yano, and H. Yamada, "*Evidence for Aharonov-Bohm effect with magnetic field completely shielded from electron wave*", Phys. Rev. Lett. **56**, no 8, pages 792-795 (1986).

[20] Y. Aharonov and A. Casher, "*Topological Quantum Effects for Neutral Particles*", Phys. Rev. Lett. **53**, no 4, pages 319-321 (1984).

[21] H. Kaiser, M. Arif, R. Berliner, R. Clothier, S. A. Werner, A. Cimmino, A. G. Klein and G. I. Opat., "*Neutron interferometry investigation of the Aharonov-Casher effect*", Physica B+C **151**, issues 1-2, pages 68-73 (1988).
https://doi.org/10.1016/0378-4363(88)90147-7

[22] A. Cimmino, G. I. Opat, A. G. Klein, H. Kaiser, S. A. Werner, M. Arif, and R. Clothier, "*Observation of the topological Aharonov-Casher phase shift by neutron interferometry*", *Phys. Rev. Lett*. **63**, no. 4, pages 380-383 (1989).

[23] M. Koenig, A. Tschetschetkin, E.M. Hankiewicz, Jairo Sinova, V. Hock, V. Daumer, et al. "*Direct Observation of the Aharonov-Casher Phase*". Phys. Rev. Lett. **96**, no. 7, 076804, (2006). [arXiv:cond-mat/0508396].

[24] D. Rohrlich, "*The Aharonov–Casher effect*", [arXiv:quant-ph/0708.3744].

[25] A. Peres, "*Unperformed experiments have no results*", Am. J. Phys. **46**, no. 7, page 745, (1978).